\documentstyle[12pt]{article}
\begin{document}
\vskip 72pt 
\centerline{\bf HOW HIGH THE TEMPERATURE OF A LIQUID} 
\centerline{\bf BE RAISED WITHOUT BOILING?}
\vskip 24pt \centerline{\bf$^{\ddag}$Mala Das, B. K. Chatterjee$^{\star}$, B. Roy and S. C. Roy$^{\dag}$}
\centerline{\it$^{\ddag}$Department of Physics, Bose Institute}
\centerline{\it$^{\ddag}$93 / 1 A. P. C. Road, Calcutta 700009, India}
\centerline{\bf$^{\ddag}$Lines to be deleted from the 
manuscript for blindfold refereeing}
\vskip 36pt 
\begin{abstract}
How high a temperature of a liquid be raised  beyond its boiling point  without 
vaporizing  (known as the limit of superheat) is an interesting subject of 
investigation. A new method of finding the limit of superheat of liquids is 
presented here. The superheated liquids are taken in the form of drops 
suspended  in a dust free gel. The temperature of the superheated liquid is 
increased very slowly from room temperature to the temperature at which the 
liquid nucleates to boiling. The nucleation is detected acoustically by a 
sensitive piezo-electric transducer, coupled to a multi channel scaler (MCS)
and the nucleation rate is observed as a function of time. The limit of superheat 
measured by the present method supersedes other measurements and theoretical 
predictions in reaching the temperature closest to the critical temperature 
of the liquids.                           
\end{abstract}
{\bf PACS} numbers: 64.60.-i, 64, 64.70.Fx, 64.60.Qb, 64.60.My
\vskip 36 pt
 
Any fluid that exists in the liquid form above its boiling temperature is said 
to be superheated. These liquids are in a metastable state in the thermodynamic 
sense and can be nucleated to form vapor by homogeneous nucleation or by the 
presence of heterogeneous nucleation sites such as gas pockets, vapor bubbles, 
solid impurities etc. or by the radiation interactions caused by charged 
particles, neutrons etc. Vapor embryos of different sizes, which are responsible 
for homogeneous nucleation, are produced at thermal equilibrium in the 
superheated liquid. The superheated state owes its existence to an energy 
barrier which causes the vapor embryo to collapse, rather than lead to 
nucleation, if it is less than a critical size. 

A liquid can not be superheated up to the critical temperature, there is a 
limit to the maximum attainable temperature for any given liquid without 
boiling. This limit is called the `limit of superheat of the liquid' (T$_{sl}$), 
where the height of the energy barrier which maintains the superheated state 
is of the order of kT and this temperature is a characteristic of any liquid. 
In addition to its importance in basic science, the knowledge of T$_{sl}$ is 
important in a number of industrial operations where a hot, nonvolatile liquid 
comes in contact with a cold volatile liquid. If the temperature of the hot 
liquid reaches to the limit of superheat of the cold liquid, explosive boiling 
would result. This explosive boiling is a potential hazard in damaging equipment 
and injure personnel in the vicinity of the blast [1]. The study of T$_{sl}$ has 
another importance since the discovery of bubble chamber by Glaser[2] and 
superheated drop detector[3]. The operation of this detector depends on the 
degree of superheat of the liquid, more the liquid is superheated more sensitive 
is the detector to lower energy radiations[4]. The minimum energy detectable by such detector 
is therefore limited by the limit of superheat of the detecting liquid. The 
limit of superheat of liquids can be estimated from the theory and can be 
measured experimentally. Theoretical calculations are performed either from 
the pure thermodynamic considerations or using the statistical mechanics. 
Very good and comprehensive reviews on homogeneous nucleation of lquid and
on the limit of superheat are available in literature[5,6,7]. 
One has to note that theoretical calculations are performed for 'pure' 
homogeneous nucleation where the chance of heterogeneous nucleation arising 
out of various interfaces with different surface energies e.g. gas-liquid, 
liquid-liquid, solid-gas etc. is completely excluded.

Experimental results reported so far are far below the critical temperature of
the liquids. One of the reasons being that observing 'pure' homogeneous 
nucleation experimentally, without any chance of heterogeneous nucleation is 
difficult to achieve. Hence, the goal is to reduce the chance of heterogeneous 
nucleation as far as possible and to use an improved method of quantitative 
detection of nucleation  to see how close one can reach experimentally 
to the predicted limit of superheat. The present experiment is designed to 
achieve this goal. Superheated sample used in this investigation is a 
homogeneous suspension of superheated drops of three liquids (R-12 : C$Cl_2F_2$,   
R-114 : ${C_2Cl_2F_4}$ and R-22 : CHCl$F_2$) in a dust free, visco-elastic, 
degassed gel medium. Suspending the superheated 
liquid in another liquid (gel) reduces the chance of heterogeneous nucleation. 
Nucleation is detected acoustically by a piezo-electric transducer[8] and the 
pulses thus received are digitized and recorded as a function of time by a multichannel 
scaler. This improved method of determining T$_{sl}$ supersedes all other 
measured values in reaching closest to the critical temperatures. Reviews on 
previous experimental techniques of measuring the limit of superheat of 
liquid have been described in detail by Avedisian[6]. As has been found 
from this literature, all previous experiments except one rely on the
qualitative observation of the nucleation  visually and therefore the present 
measurement constitutes the first quantitative measurement of T$_{sl}$ using digital 
electronics.

	The limit of superheat can be estimated either from the thermodynamic 
stability theory or from the analysis of the dynamics of formation of the 
critical sized vapor embryos (statistical mechanical theory). The superheated 
state of a liquid is a metastable state and the limit of this metastable state 
is represented on the P-V diagram by the spinodal curves. For a pure liquid, 
the spinodal curve or the thermodynamic limit of superheat is defined by states 
for which 
\begin{equation}
{\left({dP\over dV}\right)_T} = 0  
\end{equation} 
Temperley[9] calculated the value of maximum superheat temperature using van 
der Waals' equation of state. The  maximum limit of superheat of a given 
liquid can be expressed as 
\begin{equation}
t_m = {27 T_c \over32}
\end{equation}                   
where t$_m$ is the limit of superheat of the liquid . For mathematical
simplicity this has been calculated by considering the ambient pressure to be 
zero. At atmospheric pressure i.e. at P=1, t$_m$ will be slightly greater than 
the corresponding value at P=0. Other equations of state such as modified 
Bertholet equation and Redlich-Kwong equation have also been used
to calculate the limit of superheat[5]. As has been observed by Blander 
and Katz[5], experimental values of thermodynamic limit clearly exceeded 
the Van der Waals limit at least for five liquids.

	For most of the organic liquids the thermodynamic limit of superheat can be 
represented empirically[1] by 
\begin{equation}
T_{sl} = T_c [0.11(P/P_c) +0.89]
\end{equation} where T$_c$ is the critical temperature, P$_c$ is the critical 
pressure and P is the ambient pressure. 

Another method of estimating T$_{sl}$ using statistical mechanics involves 
considerations of the  rate processes of nucleation to form vapor embryos in a 
superheated liquid. This method does not yield an absolute value of T$_{sl}$ 
but it allows one to estimate the probable rate of formation of critical-sized 
vapor embryos in a superheated liquid at a given temperature. If the rate is 
very low within the time scale of the experiment, one considers no nucleation 
would occur, while if the rate is very high, then one assumes that T$_{sl}$ 
has been reached. The rate of homogeneous nucleation (J) as given approximately 
by the Volumer-Doring formula is given by[1]
\begin{equation}
J = N f{\exp{- \left({B\over kT}\right)}}
\end{equation} 
where J is the expected rate of formation of critical sized vapor embryos per 
unit volume, f is a frequency factor which in general is of the order of 
10$^{11}$ sec$^{-1}$, N is the number density of molecules in the superheated 
liquid and B is the minimum amount of energy needed  to  form  a  vapor bubble 
of critical size as given by Gibbs[10] from reversible thermodynamics is 
\begin{equation}
B = 16{\pi} {\gamma{^3}\left(T\right)}/3 {{\left({p_v} -{p_o}\right)}^2}
\end{equation}           
where $\gamma(T)$ is the liquid-vapor interfacial tension, $P_v$ is  vapor 
pressure of the superheated liquid and $P_o$ is the ambient  pressure. It is 
to be noted in this connection that which value of J is proper to calculate 
T$_{sl}$ is not defined and therefore one has to make some 'judicious choice' 
of a rate which would correspond to T$_{sl}$. A J value of 10$^6$ nucleation/cm$^3$.sec 
is often used to define the limit of superheat temperature.

It is to be noted that all the above discussions are related with the 
classical theory of nucleation. Effect of other factors like diffusion, 
viscosity and other hydrodynamical constraints are discussed by Blander 
and Katz [5]. As has been pointed out by them, contributions arising 
out of these effects in calculating T$_{sl}$ of pure liquids are not 
very significant.

The experiment is carried out with superheated liquids of R12 (b.p. -29.79
$^oC$), R114 (b.p. 3.6$^oC$) and R22 (b.p. -40.5$^oC$). The superheated 
drops are suspended in dust free, de-gassed visco-elastic gel.  The gel 
is a mixture of 'aquasonic' gel available commercially and glycerine. 
 A glass 
vial containing the superheated drops homogeneously suspended in gel is 
placed on the top of a thin layer of degassed gel taken in a beaker. The 
gel in the beaker improves the acoustic coupling between the superheated 
drops in the vial and the transducer. The beaker is placed on a piezoelectric
 transducer with a coupling gel. Some pure gel is placed on the top of the 
sample and a thermometer was inserted in the pure gel so as to avoid any 
contact with the superheated liquid sample, thus reducing the chance of 
heterogeneous nucleation from the liquid-glass interface. The nucleation in 
superheated drops is detected by the transducer, the output of the transducer
 is digitized and recorded by a multi channel scaler. The vial was wrapped 
with a heating coil covering the gel and sample. The temperature of the 
sample is increased slowly from room temperature and the count rate (dN/dt) 
is recorded in  MCS. As nucleation proceeds, the number of superheated drops 
are depleted and hence the nucleation rate is normalized with respect to the 
number of drops present. What we expect ideally is $(\frac{1}{N})$${\frac{dN}{dt}}$ 
is zero till the temperature reaches the limit of superheat where there will 
be a sudden increase in $(\frac{1}{N})$${\frac{dN}{dt}}$ (entire liquid 
nucleates) and will be no nucleation 
beyond this temperature. Considering the experimental uncertainty, one may 
observe the similar behavior as  presented in Fig. 1. The comparison of 
observed limit of superheat with other experimental results is presented in 
the table below. The reduced limit of superheat defined as $T_{sl}$/$T_c$ 
(taken in $^o$K) for these liquids is also presented in the table along with 
theoretically predicted values and other experimental results.

\vskip 10pt

\noindent\begin{tabular}{|c c|c c|c c c|c c|} \hline
 &&\multicolumn{2}{c|}{observed T$_{sl}$ $^oC$} &\multicolumn{5}{c|}{Reduced 
limit of superheat [T$_{sl}$(K)/T$_c$(K)]}\\  
\cline{5-9}
Liquid&T$_c$& & &\multicolumn{3}{|c|}{Predicted values from} 
&\multicolumn{2}{|c|}{Experiment}\\
\cline{5-7} \cline{8-9}
 & $^oK$&Present &Others&(eqn.2)&(eqn.3)&(eqn.4)&Present&Others.\\ \hline
R12  &  384.5  & 80.0 & 72.0[1] & 0.84  &  0.89  &  0.90  & 0.92  & 0.90[1]\\
&  &  & & &  & & &\\
R114  & 418.7 & 120.5 & 102.0[1] & 0.84  &  0.89  &  0.91  &  0.94  &  0.90[1]\\
&  &  &  &  &  &  & &\\
R22  &  369.0  & 57.5 & 54.0[1] & 0.84  &  0.89  &  0.89  &  0.89  &  0.89[1]\\
&  &  &  &  &  &  & & 0.89[6]\\
&  &  &  &  &  &  & &\\  \hline
\end{tabular}    
\vskip 10pt

	As could be seen from the table, the measured limit of superheat exceeded the 
predicted limit of superheat and other experimental values. It is to be noted 
in this connection that all theretical predictions are approximate as discussed 
before. Therefore the present experimental measurements indicate the need of 
improved calculation of limit of superheat. That the Van der Waals' limit is 
exceeded was reported before by Blander and Katz[5]. 	
	This table also gives an useful insight about the nucleation process. 
As can be seen from the table, for liquids with lower boiling points it is harder 
to reach closer to the critical temperature. This is quite expected as the 
chances of heterogeneous nucleation increases in case of liquids with lower 
boiling points. Whether complete elimination of heterogeneous nucleation in 
experimental measurement is possible or not is an open question. No other 
measurement have been able to reach so close to the critical temperature.  
It is to be noted in this connection that the limit of superheat of only
14 liquids out of 56 liquids studied by Blander and Katz [5] hardly exceeded 
90{\%} of the critical temperature.

	Therefore, by reducing the chances of heterogeneous nucleation by 
suspending the superheated sample in another `pure' liquid and using precise 
electronic measurement we have been able to reach closer to the critical 
temperature hitherto unattainable. Inspite of the fact that theoretical 
calculations are performed for `pure' homogeneous nucleation, they fall below
 the experimental values indicate the inadequacy of the present method of 
calculation discussed here and warrants improved calculations.  

\noindent{$^{\ddag}$[$^{\star}$Present address : Chemistry Dept. Univ. of Utah. Salt Lake 
City. USA.}]\\
\noindent$^{\ddag}$[{$^{\dag}$Author for correspondence : Fax no. 91 33 350-6790,}\\ 
email : scroy@bosemain.boseinst.ernet.in]\\ 	
\noindent$^{\ddag}${\bf Lines to be deleted from the 
manuscript for blindfold refereeing}
\vskip 24pt

\noindent{1. R. C. Reid {\it Advances in Chemical Engineering} {\bf
12}, 199 (1983).}\\
\noindent{2. D. A. Glaser {\it Phys. Rev} {\bf 8}, 665 (1952).}\\
\noindent{3. R. E. Apfel {\it US Patent} 4,143,274 (1979).}\\
\noindent{4. R.E. Apfel, S. C. Roy and Y.C. Lo {\it Phys. Rev. A.}
{\bf 31}, 3194 (1985).}\\
\noindent{5. Blander M. and Katz J. L. {\it AIChE} {\bf 21}, 833 (1975).}\\  
\noindent{6. Avedisian C. T. {\it J. Phys. Chem. Data} {\bf 14}, 695, (1985). }\\
\noindent{7. Basu D. K. and Sinha D. B. {\it Ind. J. Phys.} {\bf 42}, 
198, (1968).}\\ 
\noindent{8. Apfel R. E. and Roy S. C. {\it Rev. Sci. Inst.} {\bf 54}, 1397, 
(1983).}\\
\noindent{9. Temperley H. N. V. {\it Proc. Phys. Soc.} {\bf59}, 199, (1947).}\\ 
\noindent{10. Gibbs J. W. Translations of the Connecticut Academy III, p.108 
(1875).}\\

\end {document}